\documentclass[a4paper]{jpconf}
\usepackage{graphicx}
\usepackage{colordvi}

\def\lsi{\raise0.3ex\hbox{$<$\kern-0.75em\raise-1.1ex\hbox{$\sim$}}}
\def\gsi{\raise0.3ex\hbox{$>$\kern-0.75em\raise-1.1ex\hbox{$\sim$}}}
\newcommand{\llsim}{\mathop{\lsi}}
\newcommand{\gsim}{\mathop{\gsi}}

\newcommand{\RR}{{\kern+.25em\sf{R}\kern-.78em\sf{I} \kern+.78em\kern-.25em}}
\newcommand{\NN}{{\kern+.25em\sf{N}\kern-.78em\sf{I} \kern+.78em\kern-.25em}}
\newcommand{\CC}{{\kern+.25em\sf{C}\kern-.45em\sf{{\small{I}}} \kern+.45em\kern-.25em}}
\newcommand{\be}{\begin{equation}}
\newcommand{\ee}{\end{equation}}
\newcommand{\bea}{\begin{eqnarray}}
\newcommand{\eea}{\end{eqnarray}}
\newcommand{\nn}{\nonumber}

\begin{document}
\title{QCD in the $\delta$-Regime}

\author{W.\ Bietenholz$^{\rm a}$, N.\ Cundy$^{\rm b}$,
M.\ G\"{o}ckeler$^{\rm c}$, R.\ Horsley$^{\rm d}$, \\
Y.\ Nakamura$^{\rm e}$, D.\ Pleiter$^{\rm f}$, 
P.E.L.\ Rakow$^{\rm g}$, G.\ Schierholz$^{\rm c,h}$ \\ 
and J.M.\ Zanotti$^{\rm d}$ \ (QCDSF Collaboration)}

\address{$^{\rm a}$ Instituto de Ciencias Nucleares,
Universidad Nacional Aut\'{o}noma de M\'{e}xico, \\
\ \ \ A.P.\ 70-543, C.P.\ 04510 M\'{e}xico, Distrito Federal, Mexico \\
$^{\rm b}$ Lattice Gauge Theory Research Center, FPRD, and CTP, 
Department of Physics \& \\ \ \ \ Astronomy,
Seoul National University, Seoul, 151-747, South Korea \\
$^{\rm c}$ Institut f\"{u}r Theoretische Physik,
Universit\"{a}t Regensburg, 93040 Regensburg, Germany \\
$^{\rm d}$ School of Physics, University of Edinburgh, 
Edinburgh EH9 3JZ, United Kingdom \\
$^{\rm e}$ Center for Computational Sciences, University of Tsukuba,\\
\ \ \ Tsukuba, Ibaraki 305-8577, Japan \\
$^{\rm f}$ Deutsches Elektronen-Synchrotron DESY, 15738 Zeuthen, Germany \\
$^{\rm g}$ Theoretical Physics Division, Dept.\ of 
Mathematical Sciences, University of Liverpool, \\ 
\ \ \ Liverpool, L69 3BX, United Kingdom \\
$^{\rm h}$ Deutsches Elektronen-Synchrotron DESY, 
22603 Hamburg, Germany}

\ead{wolbi@nucleares.unam.mx}

\begin{abstract}
The $\delta$-regime of QCD is characterised by light quarks
in a small spatial box, but a large extent in (Euclidean) time. 
In this setting a specific variant of chiral
perturbation theory --- the $\delta$-expansion --- applies, based
on a quantum mechanical treatment of the quasi one-dimensional 
system. In particular, for vanishing quark masses one obtains
a residual pion mass $M_{\pi}^{R}$, 
which has been computed to the third
order in the $\delta$-expansion. A comparison with numerical
measurements of this residual mass allows 
for a new
determination of some Low Energy Constants, which appear in the
chiral Lagrangian. We first review the attempts to simulate
2-flavour QCD directly in the $\delta$-regime. This is very tedious, but
results 
compatible with the predictions for $M_{\pi}^{R}$ have been obtained. 
Then we show that an extrapolation of pion masses measured in
a larger volume towards the $\delta$-regime 
leads to good agreement with the theoretical predictions. From 
those results, we also extract a value for the 
(controversial) sub-leading Low Energy Constant $\bar l_{3}$.\\

{\tt DESY 11-019, \ Edinburgh 2011/07, \ Liverpool LTH 907}
\end{abstract}

\section{QCD and Chiral Perturbation Theory} 

\ \\
The Lagrangian of QCD is scale invariant, but its quantisation
singles out an intrinsic energy 
$\Lambda_{\rm QCD}$, which sets the scale for the hadron spectrum.
Our daily life is dominated by low energy and therefore by the
lightest quark flavours,
{\it i.e.}\ quarks with masses $m_{\rm q} \ll \Lambda_{\rm QCD}$.
In the limit of vanishing quark masses, their left- and right-handed
spinor components ($\Psi_{L}$ and $\Psi_{R}$) decouple. 
Thus the Lagrangian takes the structure
\begin{equation}
{\cal L}_{\rm QCD} = \bar \Psi_{L} D \Psi_{L} +\bar \Psi_{R} D \Psi_{R}
+ {\cal L}_{\rm gauge} \ ,
\end{equation}
where $D$ is the Dirac operator. For $N_{f}$ massless
quark flavours, this Lagrangian has the global symmetry
\be
U(N_{f})_{L} \otimes U(N_{f})_{R} = 
\underbrace{SU(N_{f})_{L} \otimes SU(N_{f})_{R}}_{\rm 
chiral~flavour~symmetry} \otimes 
\underbrace{U(1)_{V}}_{\rm baryon~number~conservation}
\otimes \underbrace{U(1)_{A}}_{\rm axial~symmetry} \ .
\ee
Here we split off the phases; the (vectorial) symmetry under 
simultaneous left- and right-handed phase rotation, $U(1)_{V}$, 
corresponds to the conservation of the baryon number. 
The remaining $U(1)_{A}$ symmetry --- for opposite 
$L$ and $R$ phase rotations --- is the axial symmetry. That is
a symmetry of the classical theory, which breaks explicitly
under quantisation, {\it i.e.}\ it is anomalous. We are interested
in the remaining {\em chiral flavour symmetry,} which (in infinite 
volume) breaks spontaneously,
\begin{equation}
SU(N_{f})_{L} \otimes SU(N_{f})_{R} \longrightarrow SU(N_{f})_{L+R} \ .
\end{equation}

{\em Chiral Perturbation Theory} ($\chi$PT) deals with an effective
Lagrangian in term of fields in the coset space of this spontaneous
symmetry breaking, $U(x) \in SU(N_{f})$ \cite{XPT}. 
Thus it captures the lightest 
degrees of freedom, which dominate low energy physics, in this case
given by $N_{f}^{2}-1$ Nambu-Goldstone bosons. The effective 
chiral Lagrangian ${\cal L}_{\rm eff}$ 
embraces all terms which are compatible with the symmetries.
This concept also extends to the case where small quark masses are
added, $m_{\rm q} \gsim 0$. Then one deals with light pseudo 
Nambu-Goldstone bosons, which are identified with the light mesons;
for $N_{f}=3$ this includes the pions, the kaons and the $\eta$-meson.

Here we consider the case $N_{f}=2$, so we only deal with the quark
flavours $u$ and $d$. We assume them to be degenerate, {\it i.e.}\
to have both the mass $m_{\rm q}$. In this case the field 
$U(x) \in SU(2)$ describes the pion triplet.
The terms in ${\cal L}_{\rm eff}$
are ordered according to an energy hierarchy, which depends on the
number of derivatives and powers of $m_{\rm q}$. 
Some of the first terms are
\bea
{\cal L}_{\rm eff} &=& \frac{1}{4} F^{2}_{\pi}\, {\rm Tr} 
[ \partial_{\mu} U^{\dagger}\partial_{\mu} U ] + \frac{1}{2}
\Sigma m_{\rm q} \, {\rm Tr} [ U + U^{\dagger} ] \nn \\
&& - \frac{1}{4} l_{1} \left( {\rm Tr} [ \partial_{\mu} U^{\dagger}
\partial_{\mu} U ] \right)^{2} - \frac{1}{4} l_{2} \left( {\rm Tr} 
[ \partial_{\mu} U^{\dagger} \partial_{\nu} U ] \right)^{2} \nn \\
&& - (l_{3} + l_{4}) \Big( \frac{\Sigma m_{\rm q}}{2 F_{\pi}} \Big)^{2}
\left( {\rm Tr} [ U + U^{\dagger} ] \right)^{2} +
l_{4} \frac{\Sigma m_{\rm q}}
{4 F_{\pi}} \, {\rm Tr} [\partial_{\mu} U^{\dagger} \partial_{\nu} U ]
\, {\rm Tr} [ U + U^{\dagger} ] + \dots
\label{Leff}
\eea
Each term comes with a coefficient, which is a free parameter
of the effective theory. These coefficients are known as
the {\em Low Energy Constants} (LECs). The leading LECs are the
pion decay constant $F_{\pi}$ (which was measured as 
$F_{\pi} = 92.4 ~ {\rm MeV}$) and the chiral condensate $\Sigma$
(the order parameter for chiral symmetry breaking). The $l_{i}$
are sub-leading LECs. In the
chiral limit $m_{\rm q}=0$, only $F_{\pi}$ occurs in the leading order.
This determines an intrinsic scale $\Lambda_{\chi} = 4 \pi F_{\pi}
\simeq 1.2 ~ {\rm GeV}$, which adopts the r\^{o}le of $\Lambda_{\rm QCD}$
in $\chi$PT. 

The LECs are of primary importance for low energy hadron physics.
To some extent they can be fixed from phenomenology. On the theoretical
side, they can only be determined from QCD as the underlying fundamental
theory. Since this refers to low energy, it is a non-perturbative
task, and therefore a challenge for lattice simulations. If one succeeds
in their determination, we arrive at a rather complete, QCD-based
formalism for low energy hadron physics.

\begin{figure}[h!]
\begin{center}
\includegraphics[width=13pc]{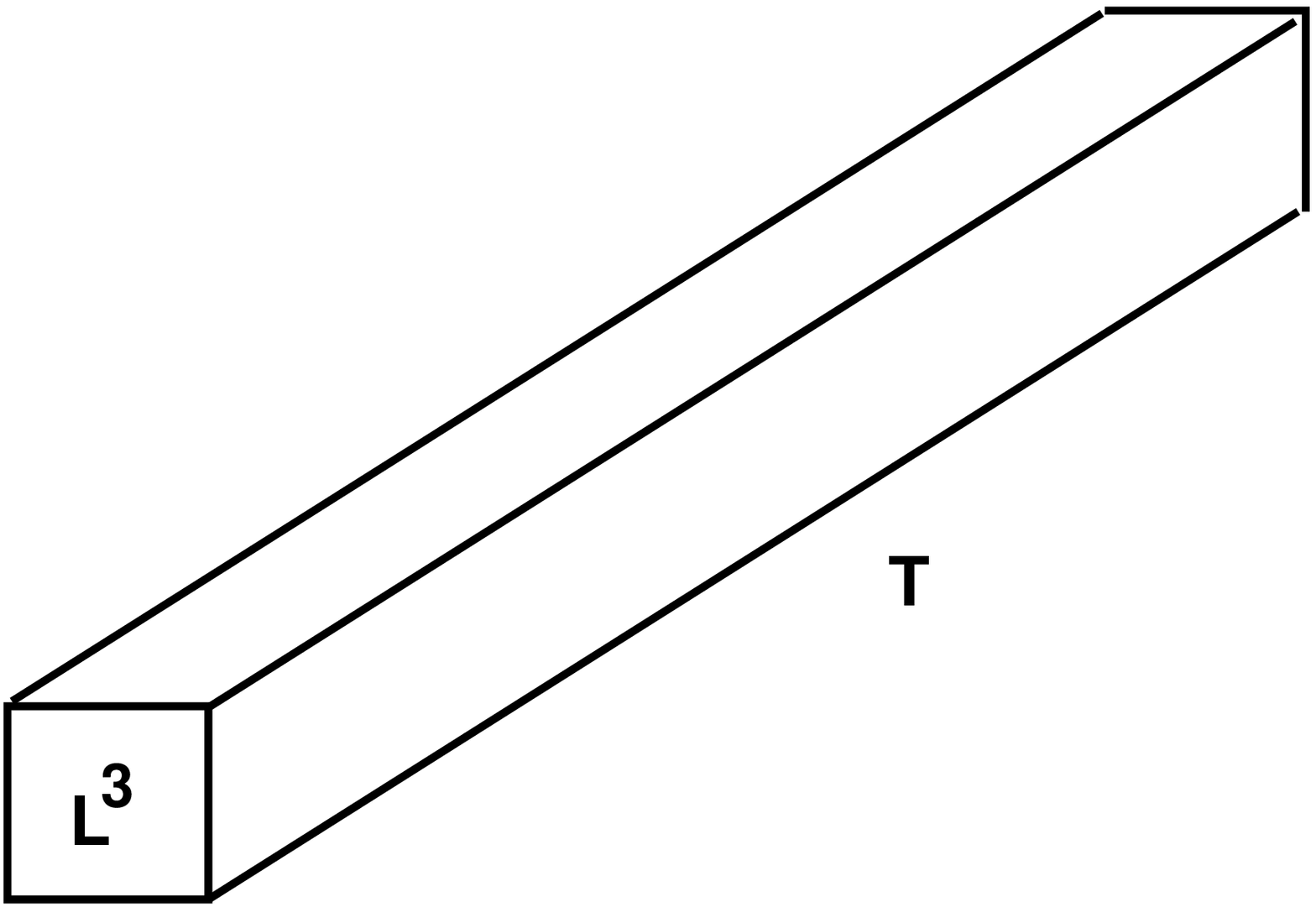} \hspace*{1cm}
\includegraphics[width=17pc]{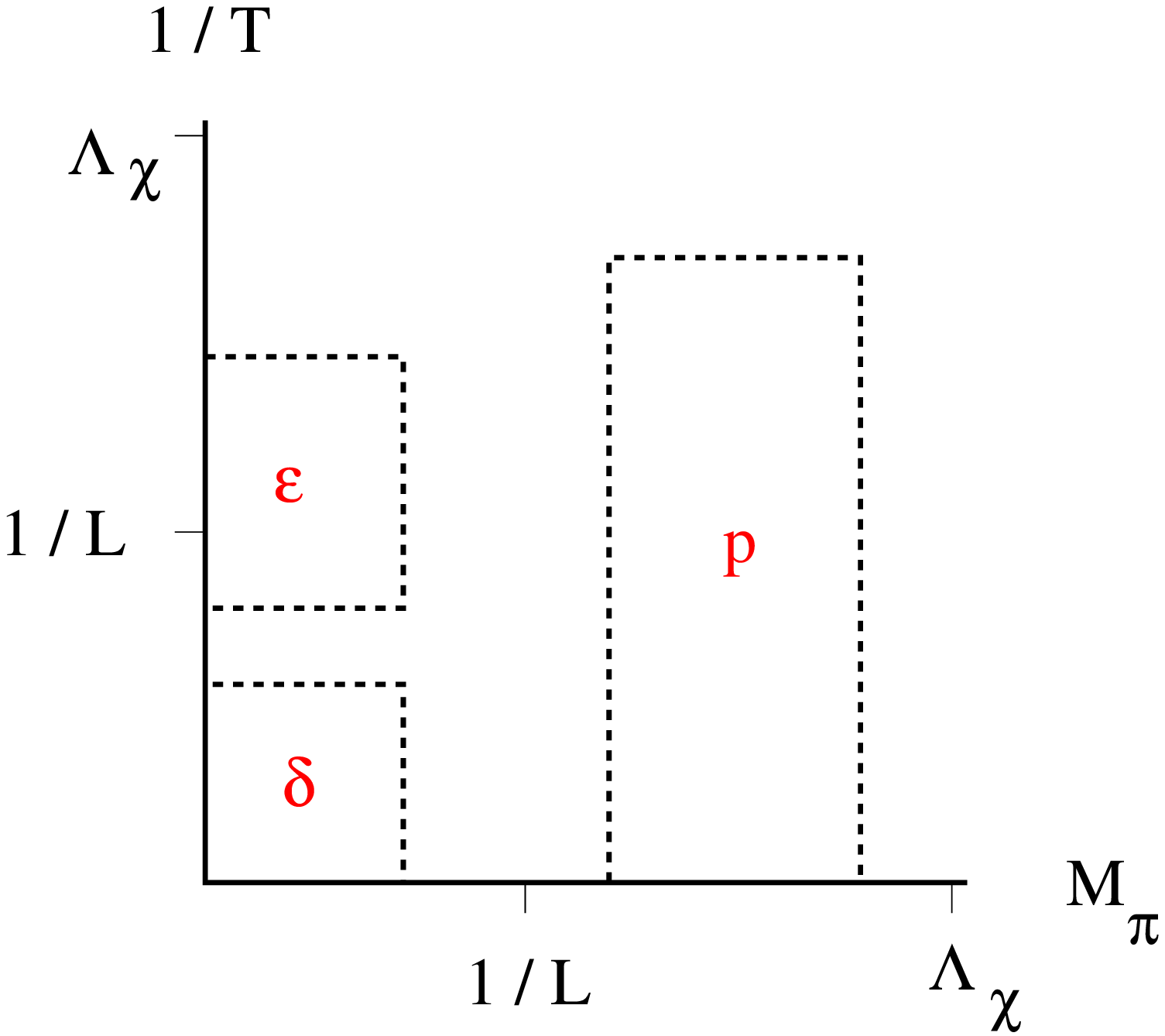}
\end{center}
\caption{\label{deltaboxmap}On the left: an illustration of
a typical shape of a $\delta$-box, {\it i.e.}\ an anisotropic
finite volume where a pion gas can be treated by the 
$\delta$-expansion. On the right: a schematic map of the applicability
domains of three different expansion rules of $\chi$PT, namely the
$p$- , the $\epsilon$- and the $\delta$-regime. The dashed lines
indicate regions where clearly one expansion holds; in the transition 
zones between these regions various expansions could work more or less.}
\end{figure}

\section{Pions in a finite volume: $p$-regime, $\epsilon$-regime and
$\delta$-regime}

\ \\
For a system of pions in a finite volume, say with
periodic boundary conditions and with a characteristic extent
$L$, the low energy expansion can be formulated in terms of the
dimensionless lowest non-zero momentum $p_{\mu}/\Lambda_{\chi} 
\sim 1/(2 F_{\pi} L)$, and of the correlation length 
$\xi = M_{\pi}^{-1}$ ($M_{\pi}$ is the pion mass).
Both $p_{\mu}$ and $M_{\pi}$
should be light compared to $\Lambda_{\chi}$. Depending on the 
size and shape of the volume, one distinguishes various 
regimes, with different counting rules for these expansion parameters: 
\begin{itemize}
\item {\underline{$p$-Regime:}} This is the standard setting with
a large volume, $L \gg \xi$, and therefore small finite size 
effects \cite{preg}.
In this case the counting is simply $O(1/L) = O(M_{\pi})$. From
the Lagrangian we can read off (on tree level) the 
Gell-Mann--Oakes--Renner relation 
\be  \label{GMOR}
M_{\pi}^{2} = \frac{\Sigma}{F_{\pi}^{2}} \, m_{\rm q} \ .
\ee
\item {\underline{$\epsilon$-Regime:}} This regime refers to a small
box in Euclidean space \cite{epsreg1}, 
say $V = L^{4}$ with $L \llsim \xi$ (where
$\xi$ is still the would-be inverse pion mass in a large volume,
with the same quark mass; in Nature $\xi \simeq 1.5 ~ {\rm fm}$).
In this regime, the $\chi$PT counting rules read
$O(1/L) = O(m_{\rm q}) = O(M_{\pi}^{2} / \Lambda_{\chi})$,
unlike the $p$-regime counting.

Experimentally the $\epsilon$-regime is not accessible, but QCD
simulations in (or at least close to)
this regime are feasible. They are of interest
in particular because the LECs determined in this regime are
the same that occur in large volume. Therefore we can
extract physical information even from an unphysical regime.
Since this can be achieved with a modest lattice size, this
method is attractive from a practical perspective \cite{eps1}.
It has been intensively explored since 2003. It is difficult to
simulate safely inside this regime, but certain
properties, which are characteristic for the $\epsilon$-regime,
have been observed. Ref.\ \cite{eps3} provides a short overview.
 
\item {\underline{$\delta$-Regime:}} Here one deals with a 
small {\em spatial} volume, but a large extent $T$ in Euclidean 
time \cite{deltareg}, say
\be
L^{3} \times T \quad , \qquad L \llsim \xi \ll T \ .
\ee
This relation for $L$ and $T$ is depicted in Fig.\ \ref{deltaboxmap}
on the left (it is exactly opposite to the setting used in studies of
QCD at finite temperature). The counting rules for the
corresponding {\em $\delta$-expansion} are
\be
\frac{1}{\Lambda_{\chi} L} = O(\delta) \quad , \qquad
\frac{M_{\pi}}{\Lambda_{\chi}} \ , \ \frac{1}{\Lambda_{\chi} T}
= O(\delta^{3}) \ .
\ee
\end{itemize}

A map of these three regimes in terms of the
pion mass and the inverse extent in Euclidean time 
(the temperature) is shown in Fig.\ \ref{deltaboxmap} on the right.

This article addresses the $\delta$-regime.
It is far less known and explored 
than the $p$- and the $\epsilon$-regime, but it shares with the
latter the exciting property that physical LECs can be extracted 
from an unphysical setting.
A further motivation for studying QCD in a ``$\delta$-box'' is that
its shape allows (approximately) for a simplified analytical treatment in
terms of 1d field theory, {\it i.e.}\ {\em quantum mechanics.}
In this case one considers a quantum rotator as described by the
1d $O(4)$ model, due to the local isomorphism between the orthogonal
group $O(4)$ and the chiral symmetry group $SU(2)_{L} \otimes 
SU(2)_{R}$. Closely related systems have applications in solid state 
physics, in particular regarding quantum anti-ferromagnets 
\cite{solid,HasNie}.

\begin{figure}[h!]
\begin{center}
\includegraphics[width=13pc,angle=270]{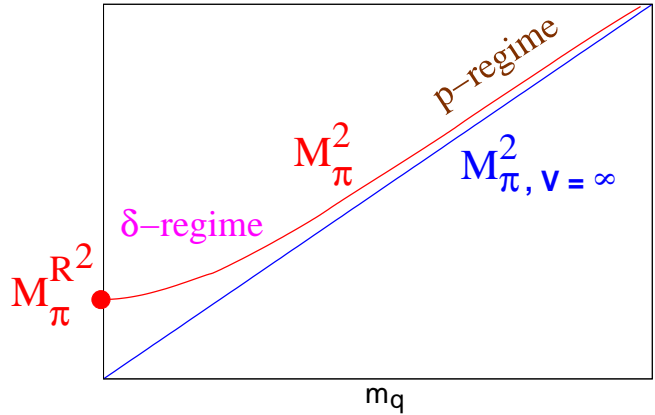}
\end{center}
\caption{\label{schema} A qualitative picture of the expected
behaviour of the pion mass squared in a $\delta$-box.
For heavy quarks and pions we approximate
the $p$-regime relation $m_{q} \propto M_{\pi}^{2}$. For light
quarks the pion mass attains a plateau, and finally (in the chiral
limit $m_{\rm q}=0$) the residual value $M_{\pi}^{R}$.}
\end{figure}

Spontaneous symmetry breaking does not occur in a finite volume.
Therefore the pions\footnote{One might argue if the term ``pion''
is adequate in the $\delta$-regime. We find it acceptable and
convenient, but readers who disagree may simply denote $M_{\pi}^{R}$
(see below) as the ``mass gap''.}
 --- {\it i.e.}\ the pseudo Nambu-Goldstone bosons ---
cannot become massless in the chiral limit $m_{\rm q}=0$, in contrast
to the infinite volume. We may consider a fixed box of a $\delta$-shape 
and vary the quark mass: a large value of $m_{\rm q}$ implies a
large pion mass, so that we enter the $p$-regime and the 
Gell-Mann--Oakes--Renner relation (\ref{GMOR}), $m_{q} \propto
M_{\pi}^{2}$, is approximated. For small 
$m_{\rm q}$ the pion mass turns into a plateau, which ends in the 
chiral limit at a {\em residual pion mass} $M_{\pi}^{R}$. This behaviour
is illustrated schematically in Fig.\ \ref{schema}.

The value of $M_{\pi}^{R}$ can be computed with
the $\delta$-expansion. The spectrum of the $O(4)$ quantum rotator
(a quantum mechanical particle on the sphere $S^{3}$) is given
by $E_{\ell} = \ell (\ell +2) / (2 \Theta)$, so the mass gap
amounts to $M_{\pi}^{R} = 3/ (2 \Theta)$. The challenge is now to 
compute the moment of inertia $\Theta$. In his seminal
paper on the $\delta$-regime, H.\ Leutwyler gave
its value to leading order (LO) as $\Theta \approx F_{\pi}^{2} L^{3}$.
Thus the residual pion mass can be written as
\be  \label{Mres}
M_{\pi}^{R} = \frac{3}{2 F_{\pi}^{2} L^{3} (1 + \Delta )} \ .
\ee
The shift $\Delta$ captures higher order corrections,
which are suppressed in powers of $1/ (F_{\pi}L)^{2}$.
They have been evaluated to next-to-leading order (NLO)
in Ref.\ \cite{HasNie}, and recently even to next-to-next-to-leading 
order (NNLO) \cite{Has}, which yields
\be  \label{Delta}
\Delta = \frac{0.4516 \dots}{F_{\pi}^{2} L^{2}} +
\frac{0.08843 \dots}{F_{\pi}^{4} L^{4}} \Big[ 1 - 0.1599 \dots
\Big( \ln (\Lambda_{1} L) + 4 \ln (\Lambda_{2} L) \Big) \Big] \ .
\ee
$\Lambda_{i}$ are scale parameters for the sub-leading LECs.
The latter are given at the scale of the physical pion mass as
\be  \label{barli}
\bar l_{i} = \ln ( \Lambda_{i} / M_{\pi}^{\rm phys} )^{2} \ .
\ee
Even more recent papers addressed again the NNLO
of the $\delta$-expansion \cite{Weier}, and the corrections
due to finite $m_{\rm q}$ \cite{Wein}.
In the following we will discuss numerical results for $M_{\pi}^{R}$.
We see that a confrontation with the analytical prediction in eqs.\ 
(\ref{Mres}), (\ref{Delta}) could enable a new 
determination of a set of LECs from first principles of QCD.

\section{Attempts to simulate QCD in the $\delta$-regime}

\ \\
The straight way to measure $M_{\pi}^{R}$ are simulations
directly {\em in} the $\delta$-regime. Since the $\delta$-box
differs from the lattice shapes in usual simulations, this requires
the special purpose generation of configurations.
Moreover, precise chirality is vital in this regime, hence
one is supposed to use a formulation of lattice quarks which
preserves chiral symmetry. Such lattice fermions are
known since the late 90ies, 
but their simulation is extremely tedious,
in particular with dynamical quarks
({\it i.e.}\ keeping track of the fermion determinant in the
generation of gauge configurations).
We anticipate that so far there are no robust results 
of simulations clearly inside the $\delta$-regime.
In this section we summarise the efforts that have been
carried out so far.

At the Symposium LATTICE2005 D.\ Hierl presented a first
attempt to simulate 2-flavour QCD in the $\delta$-regime \cite{HHHNS}.
That study used a truncated version of a chiral lattice Dirac 
operator, so a first question is if the quality of approximate 
chirality was sufficient for that purpose. Since this Dirac operator
is very complicated, that simulation was performed with a non-standard
algorithm, which probes the fermion determinant
with a stochastic estimator. The spatial volume 
was $\approx (1.2 ~ {\rm fm})^{3}$, and the results for
$M_{\pi}$ at small quark masses agreed well with the LO
of the $\delta$-expansion, {\it i.e.}\ eq.\ (\ref{Mres})
at $\Delta =0$ (and with the phenomenological value of $F_{\pi}$).

In 2007 the QCDSF Collaboration generated a new set of data, which
have not been published. They were obtained with dynamical 
overlap quarks, which are exactly chiral. This simulation used the
Hybrid Monte Carlo algorithm ({\it i.e.}\ the reliable standard
algorithm). The lattice had a modest size of $8^{3} \times 16$ sites, 
and the spatial box length was again $L  \approx 1.2 ~ {\rm fm}$.
At first sight the results seemed to look
fine: for decreasing $m_{\rm q}$ we saw a transition from a
Gell-Mann--Oakes--Renner type behaviour towards a plateau. Its value
agreed with the chirally extrapolated value of Ref.\ \cite{HHHNS},
and therefore also with the LO of eq.\ (\ref{Mres}).
  

Unfortunately, this is {\em not} the end of the story. If we proceed
to the NLO correction, {\it i.e.}\ if we include the first term 
of $\Delta$ given in eq.\ (\ref{Delta}), the predicted value for
$M_{\pi}^{R}$ decreases drastically in this small volume --- from 
$782.5 ~{\rm MeV}$ down to $321.7 ~{\rm MeV}$ --- and the agreement
with the above data is gone. Considering this dramatic effect of
the NLO correction one might worry that the $\delta$-expansion
could converge only very slowly in this small box, 
and such simulations are not instructive at all. However, 
adding also the NNLO correction 
alters the NLO result only a little --- to 
$(336.3 \pm 7.6) ~{\rm MeV}$ --- so it is reasonable to assume 
the $\delta$-expansion to be already well converged.\footnote{To 
obtain this theoretically predicted value, we inserted the LECs as
far as they are known. In particular the sub-leading LECs 
$\bar l_{1} = -0.4 \pm 0.6$ and $\bar l_{2} = 4.1 \pm 0.1$ 
are taken from Ref.\ \cite{CGL}, along with their 
uncertainties, which imply the uncertainty in the NNLO value of 
$M_{\pi}^{R}$.\label{lifoot}}

\begin{figure}[h!]
\begin{center}
\vspace*{-6mm}
\includegraphics[width=17.9pc,angle=270]{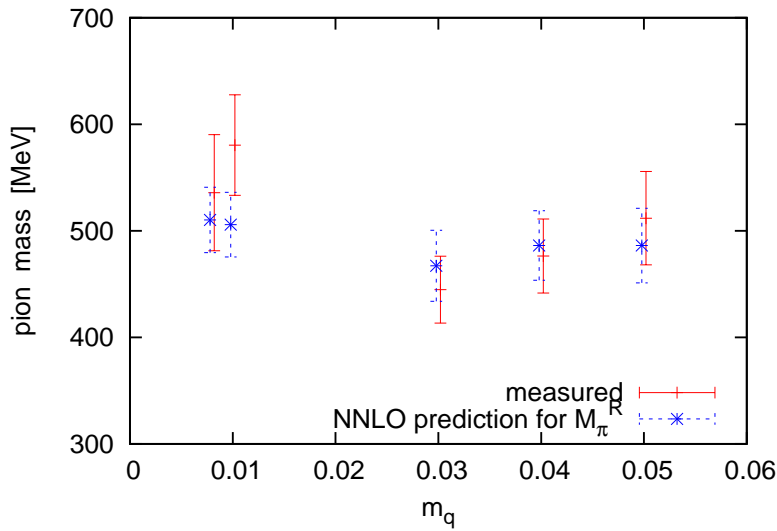}
\end{center}
\vspace*{-4mm}
\caption{An illustration of our results given in Table \ref{run2dat}:
we show the numerically measured pion mass 
and the theoretical residual mass $M_{\pi}^{R}$, as predicted by the
NNLO $\delta$-expansion. (The latter actually refer to the chiral
limit, but its value depends on $L$ and therefore
on the lattice spacing $a$, which varies for the simulations
at different $m_{\rm q}$.) We observe good agreement between the
measured and the predicted values.}
\label{deltadat}
\vspace*{-5mm}
\end{figure}

Thus simulations in this small box {\em can} be useful
in view of a confrontation with the analytical predictions for
$M_{\pi}^{R}$. In fact a second sequence of runs by the QCDSF
Collaboration (performed in 2008) yielded data much closer to
the NNLO prediction. They involved 5 quark masses; the 
results for the physical lattice spacing $a$ and the pion mass 
are given in Table \ref{run2dat}. 
Since $a$ varies for the different values of $m_{\rm q}$,
the box length $L=8a$ and the prediction for $M_{\pi}^{R}$ vary 
as well. 
In Fig.\ \ref{deltadat} 
we compare the numerically measured pion masses and the corresponding
NNLO $\delta$-expansion results. They are compatible, in contrast to 
the earlier data, which were probably not well thermalised. 
In these new runs thermalisation is accomplished, but the ratio
$T/L=2$ is still modest. We therefore proceeded to a 
$8^{3}\times 32$ lattice, where our runs are ongoing. 
Still, the quality of agreement with the $\delta$-expansion
that we observed already on the $8^{3}\times 16$ lattice is impressive. 
\begin{table}
\begin{center}
\begin{tabular}{|c||c|c|c|}
\hline
$m_{\rm q}$ & $a$ [fm] & $M_{\pi}$ & $M_{\pi}^{R}$ (NNLO) \\
\hline
0.008 & 0.102(9)& $(535.8 \pm 54.5)$ MeV & $(510.3 \pm 30.7)$ MeV \\
\hline
0.01 & 0.104(8) & $(580.5 \pm 47.2)$ MeV & $(505.8 \pm 30.3)$ MeV \\
\hline
0.03 & 0.114(8) & $(444.8 \pm 31.4)$ MeV & $(476.2 \pm 33.4)$ MeV \\
\hline
0.04 & 0.111(8) & $(476.4 \pm 34.7)$ MeV & $(486.2 \pm 32.6)$ MeV \\
\hline
0.05 & 0.111(9) & $(511.9 \pm 43.9)$ MeV & $(486.2 \pm 35.0)$ MeV \\
\hline
\end{tabular}
\end{center}
\caption{The results by QCDSF Collaboration of the year 2008, on 
a $8^{3} \times 16$ lattice with 5 values of the mass $m_{\rm q}$
for the two degenerate dynamical overlap quark flavours.
We display the measured values of the physical lattice spacing $a$ 
and the pion mass, as well as the
NNLO prediction for the residual pion mass $M_{\pi}^{R}$.
(The errors capture the uncertainty in $a$, in the pion mass in
lattice units, and in the LECs $\bar l_{1}$ and $\bar l_{2}$,
cf.\ eq.\ (\ref{l1234}).)}
\label{run2dat}
\vspace{-5mm}
\end{table}

\section{Residual pion mass by an extrapolation from the
 $p$-regime}

\ \\
In this section we proceed to a different approach.
It is based on simulation results in the $p$-regime (up to the
transition zone), 
which are then extrapolated
towards the $\delta$-regime. Details of this study are given in
Ref.\ \cite{QCDSFdelta}. Also in this framework we consider it essential
to use dynamical quarks, but we do not insist on exact chiral symmetry
in the $p$-regime. Hence we used Wilson 
fermions, which is an established standard lattice fermion formulation,
in a form which corrects $O(a)$ lattice artifacts. Thus the simulation 
was much faster, and we could tackle much larger lattices than those
mentioned in the Section 3; our data reported below were
obtained on three lattice sizes:
$ 24^{3} \times 48 \ , \ 32^{3} \times 64 $ and $40^{3} \times 64 \ .$
On the other hand, this lattice regularisation breaks 
the chiral symmetry explicitly, so that additive mass 
renormalisation sets in. Nevertheless we were able
to attain very light pion masses.

For the gauge part we used the standard plaquette lattice action.
Our simulations were carried out at two values for the strong
gauge coupling $g_{\rm s}$, respectively the parameter 
$\beta = 6 / g_{\rm s}^{2}$. We determined the physical
lattice spacing $a$ from the measured nucleon mass,
which revealed that we were dealing with fine lattices,
\be
\beta = 5.29 \ \to \ a \simeq 0.075 ~ {\rm fm} \quad ,
\quad 
\beta = 5.4 \ \to \ a \simeq 0.067 ~ {\rm fm} \quad .
\ee
Thus the spatial size was in the range
$L \simeq 1.6 ~{\rm fm} \dots  3.0 ~{\rm fm} .$
Due to the additive mass renormalisation, we could not refer to the 
bare quark mass anymore. We measured the current
quark mass by means of the PCAC relation,
\be
m_{\rm q} = \frac{\langle 
\partial_{4} A_{4} (\vec 0 , x_{4}) P(0) \rangle}
{\langle P (\vec 0 , x_{4}) P(0) \rangle} \ ,
\ee
where $P$ is the pseudoscalar density, and $A_{\mu}$ is the
axial current. 
We observed practically no finite size effects on $a$ 
and on $m_{\rm q}$, but we did see a striking $L$-dependence of 
$M_{\pi}$, as expected. These quantities were found in the range
\be
m_{\rm q} = 3.60 ~ {\rm MeV} \dots 231 ~ {\rm MeV} \quad , \quad
M_{\pi} = 174 ~ {\rm MeV} \dots 1.52 ~ {\rm GeV} \quad  
, \quad M_{\pi} L = 2.7 \dots 9.7 \ .
\ee
The latter confirms that our data range from the deep $p$-regime
to the transition zone. A few missing data points
have been completed with a (lengthy) formula for exponentially
suppressed finite size effects \cite{ML,CDH} 
within the $p$-regime, which holds up to $O(p^{4})$. 
This formula involves the renormalised sub-leading LECs
$\bar l_{i}, \ i=1 \dots 4$, see eqs.\ (\ref{Leff}) and (\ref{barli}).
Well established phenomenological values 
were composed in Ref.\ \cite{CGL}
($\bar l_{1}, \ \bar l_{2}$ were anticipated in footnote \ref{lifoot}),
\be  \label{l1234}
\bar l_{1} = -0.4 \pm 0.6 \quad , \quad \bar l_{2} = 4.1 \pm 0.1
\quad , \quad \bar l_{3} = 2.9 \pm 2.4 
\quad , \quad \bar l_{4} = 4.4 \pm 0.4 \quad .
\ee
They were estimated from
$\pi \pi$ scattering data, and in particular the
$\bar l_{4}$ value is based on the scalar pion form factor.
The dark horse in this context is $\bar l_{3}$:
we replaced the above value by  $\bar l_{3} \approx 4.2$,
which we obtained in our study, see below.

\begin{figure}[h!]
\begin{center}
\includegraphics[width=17.3pc,angle=0]{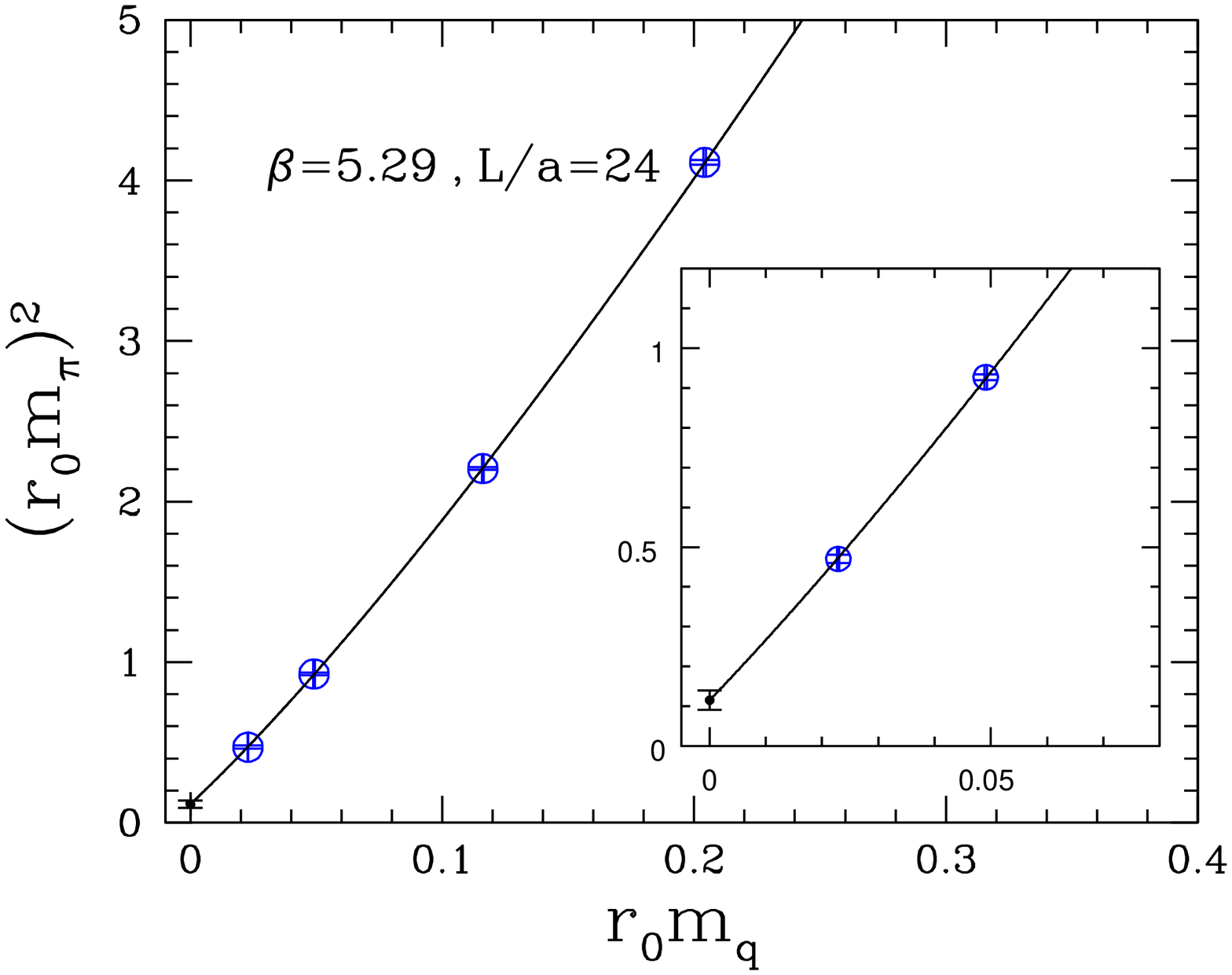}
\includegraphics[width=17.3pc,angle=0]{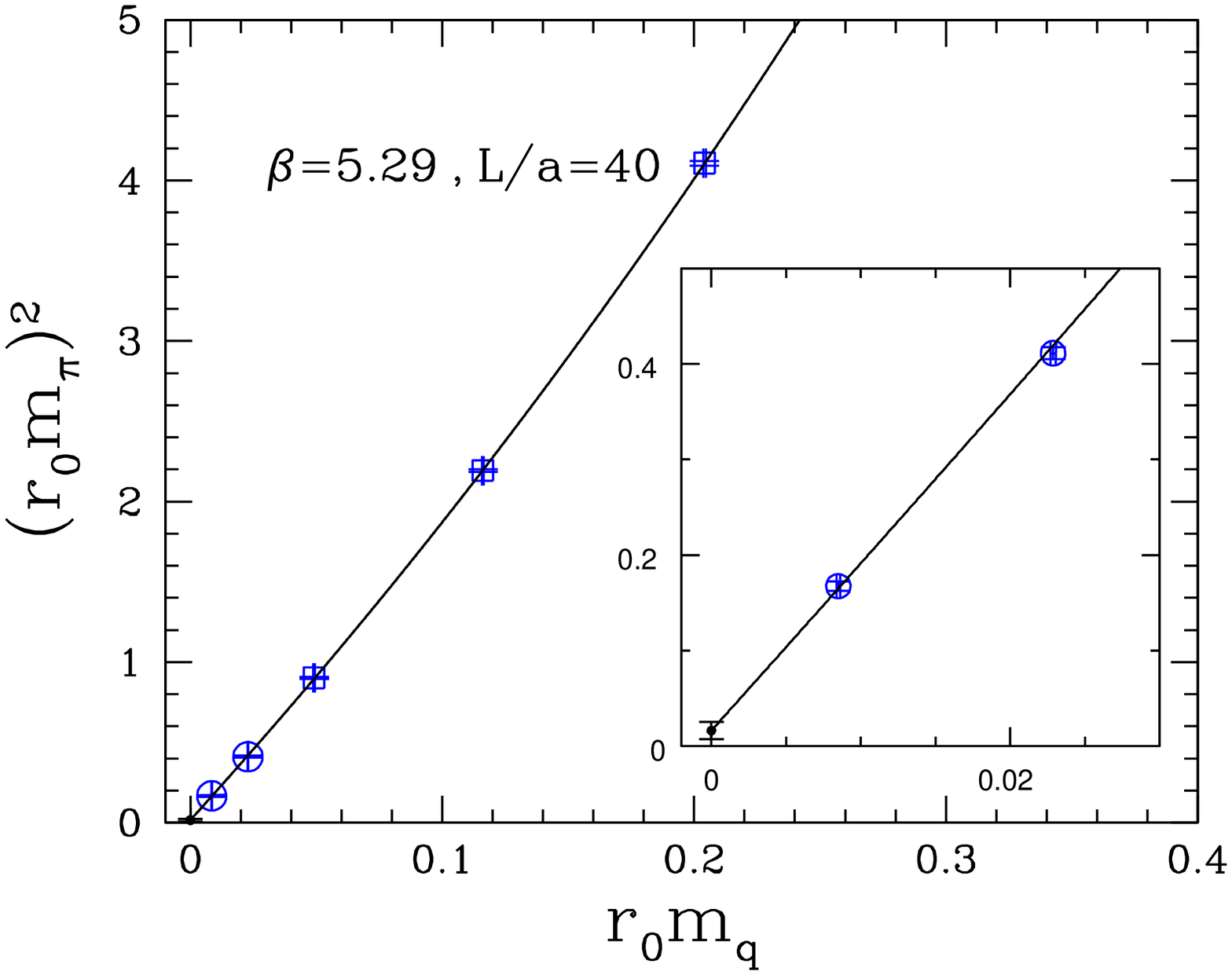}
\end{center}
\vspace*{-2mm}
\caption{Our chiral extrapolation referring to
the lattice spacing $a \simeq 0.075 ~ {\rm fm}$ (corresponding
to $\beta =5.29$) with $L \simeq 1.8 ~ {\rm fm}$ (on the left)
and $L \simeq 3.0 ~ {\rm fm}$ (on the right).
The data points show the measured pion mass 
squared against the quark mass in the $p$-regime (the Sommer
scale parameter $r_{0}= 0.467 ~ {\rm fm}$ is employed to
convert them into dimensionless units). The curve is the fit
according to eq.\ (\ref{fitfun}). It ends in the chiral limit,
where we illustrate the extrapolated value for $M_{\pi}^{R}$ and its
error.}
\label{b529}
\end{figure}

\begin{figure}[h!]
\begin{center}
\vspace*{-1.3cm}
\includegraphics[width=17.3pc,angle=0]{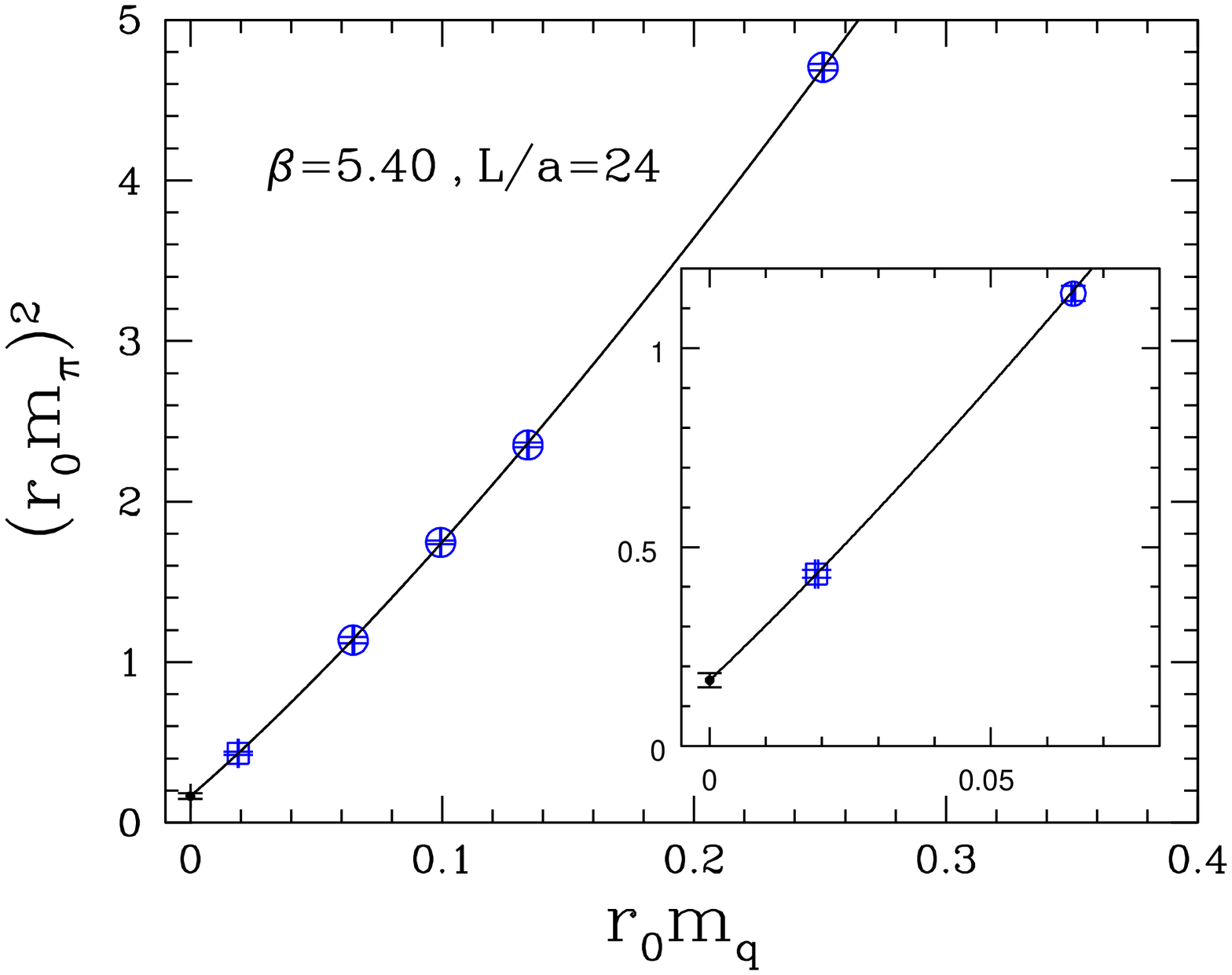}
\includegraphics[width=17.3pc,angle=0]{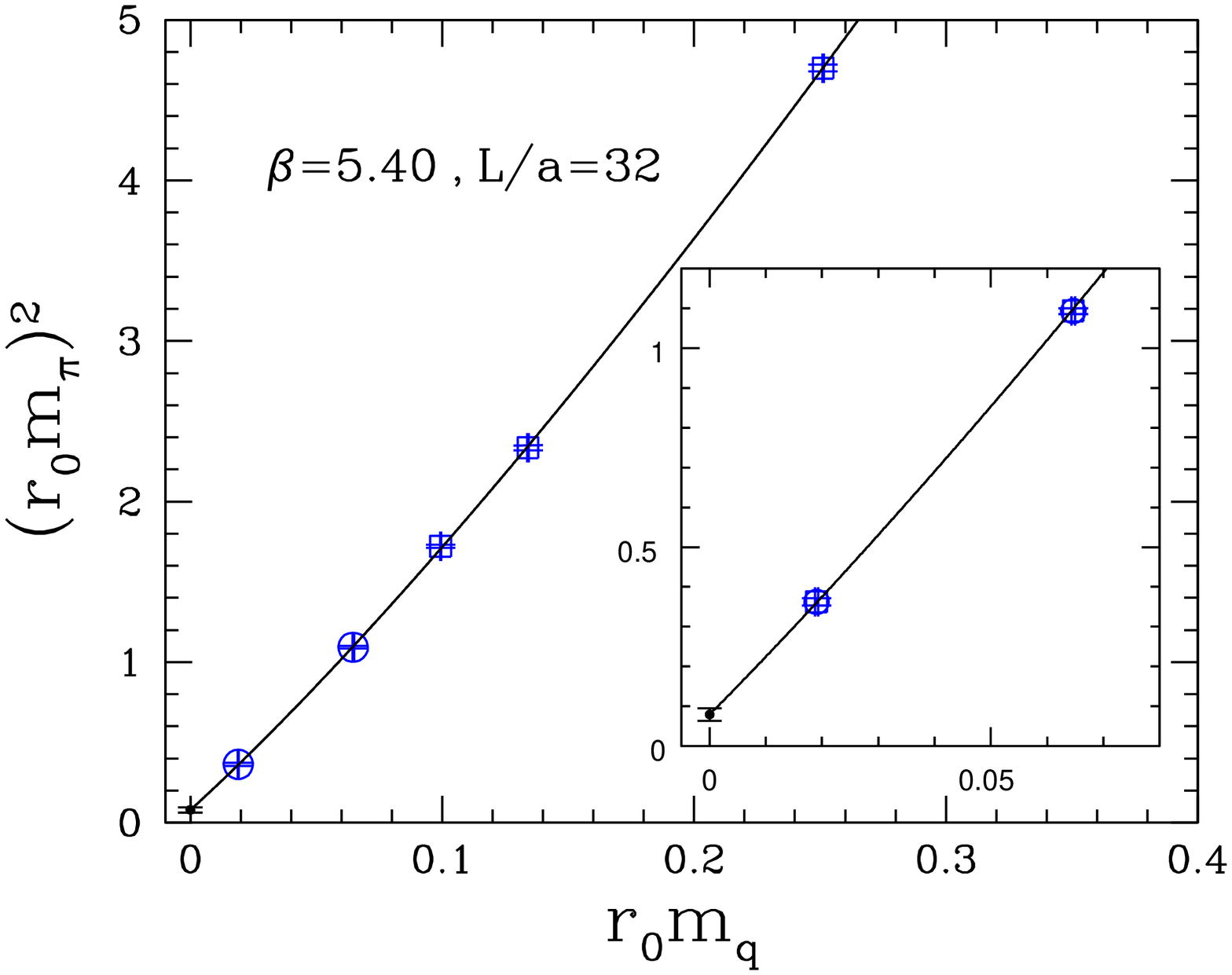}
\end{center}
\vspace*{-2mm}
\caption{The same as Fig.\ \ref{b529}, but now for the finer
lattices with $a \simeq 0.067 ~ {\rm fm}$ (corresponding
to $\beta =5.4$), such that $L \simeq 1.6 ~ {\rm fm}$ (on the left)
and $L \simeq 2.1 ~ {\rm fm}$ (on the right).}
\vspace*{-2mm}
\label{b540}
\end{figure}

Our measured and interpolated data are given in
Ref.\ \cite{QCDSFdelta}. We extrapolated them towards 
the $\delta$-regime and extracted in particular a value for 
$M_{\pi}^{R}$ based on the relatively simple 
chiral extrapolation formula
\be  \label{fitfun}
M_{\pi} (L)^{2} = M_{\pi}^{R~2} + C_{1} m_{\rm q} 
\left[ 1 + C_{2} m_{\rm q} \ln (C_{3} m_{\rm q}) \right] \ ,
\ee
which interpolates between the $O(p^{4})$ correction formula
(for large $L M_{\pi}$) and the chiral limit \cite{CGL}.
 The $C_{i}$ and $M_{\pi}^{R}$ are treated as free parameters
to be fixed by the fit. Our data and the fits for 
$\beta = 5.29$ and for $\beta = 5.4$ are shown in Figs.\
\ref{b529} and \ref{b540}, respectively. These plots also
illustrate the extrapolation result for $M_{\pi}^{R}$
in the chiral limit. 

\begin{figure}[h!]
\begin{center}
\includegraphics[width=19pc,angle=270]{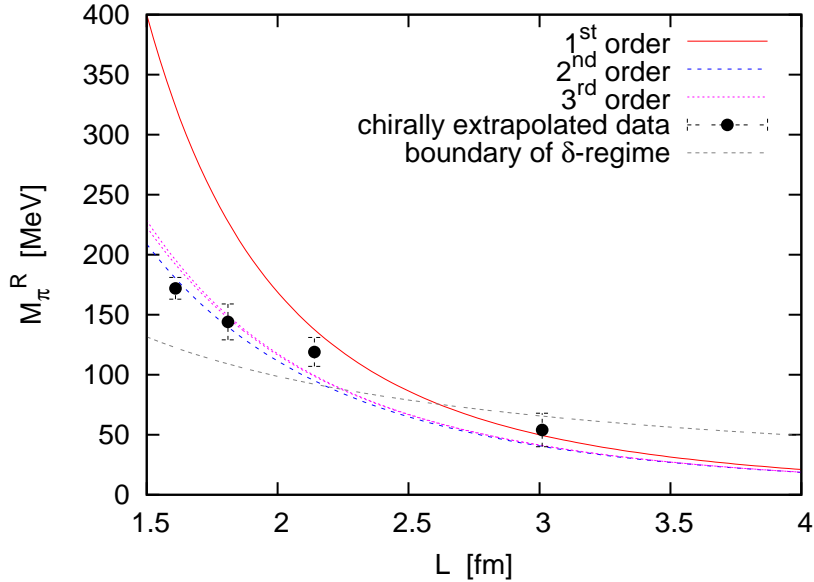}
\end{center}
\caption{A comparison of our extrapolated values for the
residual pion mass $M_{\pi}^{R}$ with the predictions
by the $\delta$-expansion (LO, NLO, NNLO), as functions of 
the spatial box size $L$. We observe a remarkably good agreement.
In our largest volume, the extrapolated $M_{\pi}^{R}$ is located
below the physical pion mass, and in the region $M_{\pi}^{R} L < 1$,
which can be viewed as the domain of the $\delta$-regime.}
\label{finale}
\end{figure}

Our main result is shown in Fig.\ \ref{finale}. It compares
the extrapolated values for $M_{\pi}^{R}$ with the predictions
based on the $\delta$-expansion to LO, NLO and NNLO, as a function
of $L$. The latter two predictions are very close to each other
for the volumes under consideration, so we can again assume the 
expansion to be well converged. 

The extrapolation results for $M_{\pi}^{R}$
reach down to values even below the physical pion mass.
The plot also shows the line where the product $M_{\pi}L$
decreases below 1; this can be roughly considered as
the boundary of the $\delta$-regime. Our extrapolations
lead close to this boundary, and in one case {\em into} the
$\delta$-regime.\footnote{In that case, the extrapolation
should actually turn into a plateau at tiny $m_{\rm q}$,
but this would hardly change the result.}

In particular this plot shows that the extrapolated masses
$M_{\pi}^{R}$ match this curve 
remarkably well. It would not have been obvious to predict this
feature, because our data were obtained in a regime where the
basis of the $\delta$-expansion (frozen spatial degrees of freedom) 
does not hold. 
This suggests that the extrapolation formula (\ref{fitfun}) applies
well in a sizable domain. 

\section{Evaluation of Low Energy Constants}

\ \\
By fitting our pion mass data according to eq.\ (\ref{fitfun}),
we obtained results for the four free fitting parameters.
In Section 4 we discussed the results for $M_{\pi}^{R}$ that
we obtained in this way. Moreover, the results for $C_{1}$,
$C_{2}$ and $C_{3}$ can be used to evaluate the notorious
sub-leading LEC $\bar l_{3}$ \cite{QCDSFdelta}. This is how
we obtained our value that we anticipated in Section 4,
\be
\bar l_{3} = 32 \pi^{2} F_{\pi}^{2} \, \frac{C_{2}}{C_{1}} \, 
\ln \frac{C_{1}}{C_{3} M_{\pi}^{{\rm phys}~2}} 
= 4.2 \pm 0.2 \ .
\ee
(The error given here emerges from the fits, the additional systematic
error would be hard to estimate).
In view of other results in the literature, this value is in 
the upper region. An overview has been presented in Ref.\
\cite{Latdat}, in particular in Table 11 and Figure 9.
That overview estimates the world average as $\bar l_{3} = 3.3 (7)$. 

In the previous considerations of Sections 4 and 5, we 
inserted the phenomenological value of the pion decay constant,
$F_{\pi} = 92.4 ~{\rm MeV}$. Alternatively, we could also
treat $F_{\pi}$ as a free parameter to be determined by the fits.
In particular, matching $M_{\pi}^{R}$ this yields 
in the chiral limit \cite{QCDSFdelta}
\be
\left. F_{\pi}^{\rm numerical} \right\vert_{m_{\rm q}=0} = 78^{+14}_{-10}
~{\rm MeV} \ ,
\ee
which seems a bit low. However, effective field theory considerations
suggest that the value of $F_{\pi}$ in the chiral limit
should indeed be below the physical value; Ref.\ \cite{Latdat} 
estimates $F_{\pi} \vert_{m_{\rm q}=0} \approx 86 ~ {\rm MeV}$.

\section{Conclusions}

\ \\
The $\delta$-regime refers to a system of pions in a finite
box, typically of the shape $L^{3} \times T$ with 
$L \llsim M_{\pi}^{-1} \ll T$. It can be treated by Chiral 
Perturbation Theory with suitable counting rules, 
the $\delta$-expansion. In particular this predicts
the residual pion mass in the chiral limit, $M_{\pi}^{R}$,
which has been computed recently to NNLO \cite{Has}.

First attempts to simulate 2-flavour QCD in the $\delta$-regime
were confronted with technical difficulties, in particular
thermalisation problems. Nevertheless we obtained good results for the
pion mass with light quarks on a small lattice of size 
$8^{3} \times 16$, with $L \simeq (0.82 \dots 0.91) ~{\rm fm}$.

In another pilot study we measured pion masses in the $p$-regime
(up to the transition region) and extrapolated them towards the
$\delta$-regime. This yields numerical results for $M_{\pi}^{R}$,
which agree remarkably well with the predictions by the
$\delta$-expansion. That expansion is based on assumptions, 
which do not hold in the regime where the data were obtained.
Therefore it would not have been obvious to predict the observed 
agreement. This comparison can be viewed as a numerical
experiment, which led to an interesting observation.

Our pion mass fits from the $p$- towards the $\delta$-regime 
also fix some further constants, which allow
for a new determination of the (mysterious) sub-leading
Low Energy Constant $\bar l_{3}$; we obtained $\bar l_{3} = 4.2(2)$.
This is somewhat above the average of the phenomenological
and numerical estimates in the literature. The constants
$\bar l_{1}, \dots , \bar l_{4}$ are relevant in the effective 
description of $\pi \pi$ scattering.
Finally we also considered the option to treat $F_{\pi}$
as a free parameter. In the chiral limit the fits lead to a
value somewhat below the phenomenological $F_{\pi}$.

Robust numerical results, measured manifestly inside the $\delta$-regime,
are still outstanding.\\

\vspace*{-1mm}
\noindent
{\bf Acknowledgements :} W.B.\ thanks for kind hospitality
at the University of Bern, where this talk was written up.
The simulations for this project were performed on the SGI Altix 4700
at LRZ (Munich), the IBM BlueGeneL and BlueGeneP at NIC (J\"{u}lich),
the BlueGeneL at EPCC (Edinburgh), and the apeNEXT at NIC/DESY
(Zeuthen). We also thank for financial support by the EU Integrated
Infrastructure Initiative ``HadronPhysics2'' and by the DFG under
contract SFB/TR 55 (Hadron Physics from Lattice QCD). J.M.Z.\ is supported
through the UKs STFC Advanced Fellowship Program under contract
ST/F009658/1.

\end{document}